\documentclass[prl,twocolumn,amsmath,amssymb,superscriptaddress,showpacs]{revtex4-1}

\usepackage{natbib}
\usepackage{graphicx}
\usepackage[usenames]{color}

\newcommand{\fig}[1]{Fig.~\ref{#1}}
\newcommand{\be}[1]{\begin{equation}\label{#1}}
\newcommand{\ee}{\end{equation}}

\begin{document}

\title{Frequency Resolved Optical Gating for Time-Resolving Intra-Atomic Knock-Out in Double Ionization with Attosecond Pulses}
\author{H. Price$^{1}$, A. Staudte$^{2}$, P. B. Corkum$^{2}$ and  A. Emmanouilidou$^{3,}$}

\affiliation{Department of Physics and Astronomy, University College London, Gower Street, London WC1E 6BT, United Kingdom\\
$^2$Joint Laboratory for Attosecond Science, University of Ottawa and National Research Council, 100 Sussex Drive, Ottawa, Ontario, Canada K1A 0R6 \\
$^3$Chemistry Department, University of Massachusetts at Amherst, Amherst, Massachusetts, 01003, U.S.A.}

\begin{abstract}
We develop the two-electron attosecond streak camera under realistic conditions using a quasi-classical model. We assume extreme ultra-violet (XUV) attosecond pulses with a full width at half maximum (FWHM) of 24 attoseconds, centered at 120~eV and a streaking infrared laser field of 1600~nm, and intensity of $1.8 \times 10^{12}$ W/cm$^2$. The proposed method is shown to be capable to time resolve intra-atomic collisions in double ionization.
\end{abstract}
\maketitle

Time-resolving correlated electron processes is one of the driving forces behind the large scale effort to push the frontiers of attosecond science. Attosecond science offers time resolution through XUV-pulses. However, pump-probe experiments using attosecond pulses are technically very challenging. Hence, the streaking of photo-electrons with an infrared (IR) laser field has become a successful technique for bringing time resolution to photo-ionization. The paradigmatic attosecond streak camera \cite{Drescher2001Science,Itatani2002PRL}, and its development into FROG CRAB (Frequency Resolved Optical Gating for Complete Reconstruction of Attosecond Bursts) \cite{Quere2005PRA} originally aiming to characterize attosecond extreme ultra-violet (XUV)-pulses, have been the underlying concept for studies resolving delayed time emission from atoms \cite{Uiberacker2007Nature,Eckle2008Science, Schultze2010Science, KlunderPRL2011,Taylor} and solids \cite{Cavalieri2007Nature}. 

To study the electron correlation in single photon double ionization we recently formulated the two-electron streak camera \cite{Emmanouilidou2010NJP}. Specifically, it was shown that the intra-atomic  knock-out process \cite{Knapp2004PRL} can be associated with a delay, i.e.  a time between photo-absorption and ejection of two electrons in the continuum. The delay is encoded in the inter-electronic angle of escape  as a function of the phase  of the IR laser field. In addition,  the two-electron streak camera  can time-resolve delays
corresponding to different energy sharings, between the two electrons, and to different ionization mechanisms \cite{Emmanouilidou2011NJP}. However, previous work considered only discrete photon energies and instantaneous photo absorption \cite{Emmanouilidou2010NJP,Emmanouilidou2011NJP}.

Here, we remove this severe limitation and extend the two-electron streak camera to realistic attosecond pulses. By resolving the bandwidth of an XUV-pulse in the sum-energy of two emitted electrons we construct the two-electron equivalent of FROG to obtain a complete picture of the single photon double ionization process. Specifically, in FROG \cite{Trebino} one extracts from  a two-dimensional data set (FROG-trace) the complete characteristics of an optical pulse. In a similar manner in FROG-CRAB \cite{Quere2005PRA}  one  retrieves the spectral phases and amplitudes of an attosecond pulse. Here, we assume a fully characterized, transform limited XUV attosecond pulse, to obtain information about the delay of two-electron emission after absorption of a photon from the attosecond pulse. We use as  FROG-like trace for double ionization the inter-electronic angle of escape as a function of the phase of the IR laser field. A simple algorithm is devised for extracting the intra-atomic two-electron delays for different excess energies.

We build on our classical trajectory Monte-Carlo \cite{Abrines1966PPS} simulation of the classical He$^*$(1s2s) model system, for details see \cite{Emmanouilidou2010NJP,Emmanouilidou2011NJP,Emmanouilidou2007PRA1, Emmanouilidou2008PRL,Emmanouilidou2006JPB}.
  We choose He$^*$(1s2s) as a prototype system to clearly formulate the concept of streaking
two-electron dynamics while avoiding the unnecessary complexity of many-electron systems.
However, the scheme is not system specific. The same scheme could time-resolve, for instance,
the collision between 1s and 2s electrons in the ground state of Li \cite{Emmanouilidou2007PRA1,Emmanouilidou2008PRL,Emmanouilidou2006JPB}.  Atomic units (a.u.) are
used throughout this work except where otherwise indicated. 

\begin{figure}
  \includegraphics[width=0.5\textwidth]{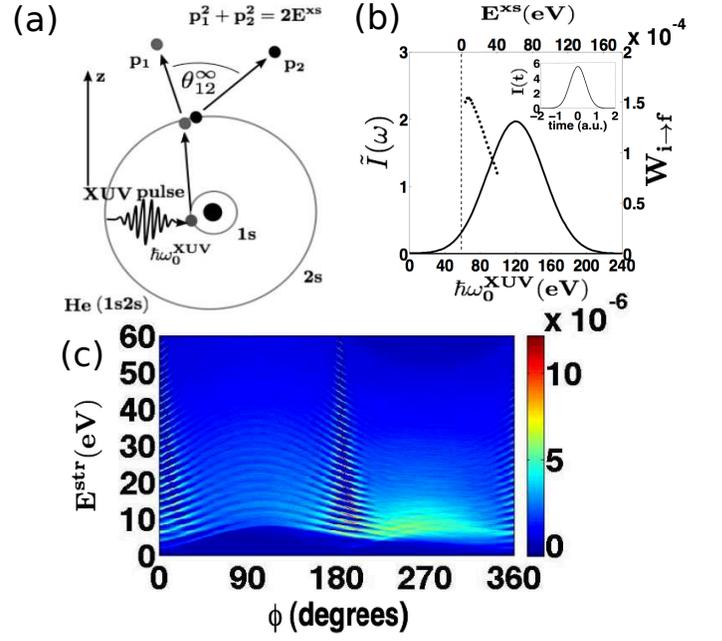}           
 \caption{(a) Sketch of the intra-atomic knock-out mechanism to be studied with the two-electron streak camera. The streaking field causes a decrease in $\mathrm{\theta_{12}^{\infty}}$when the photo-electron
is launched along the +$\mathrm{\hat{z}}$ direction, since adding $\mathrm{\Delta p_{IR}}$ to each of the
electron momenta results in  $\mathrm{\theta_{12}<\theta_{12}^{\infty}}$. (b) Spectral intensity of the XUV-pulse scaled by $\mathrm{(F_{0}^{XUV})^2}$. Dotted curve:  $\mathrm{W_{i\rightarrow f}}$ in the presence of the XUV and IR field averaged over all $\phi's$. (c) Observable $\mathrm{E^{str}}$ total electron energy as a function of $\mathrm{\phi}$ considering excess energies from 4 eV to 60 eV in steps of 2 eV and double ionization events corresponding to launching of the photo-electron (1s) in the $\pm \mathrm{\hat{z}}$ direction for $\mathrm{\phi}$ ranging from 0$^{\circ}$ to 180$^{\circ}$. To illustrate the difference between launching of the photo-electron 
in the + $\mathrm{\hat{z}}$ versus the  -$\mathrm{\hat{z}}$ direction we plot the $\mathrm{E^{str}}$ corresponding to + $\mathrm{\hat{z}}$ for $\mathrm{\phi}$ ranging from 0$^{\circ}$ to 180$^{\circ}$ and the $\mathrm{E^{str}}$ corresponding to - $\mathrm{\hat{z}}$ for $\mathrm{\phi}$ ranging from 180$^{\circ}$ to 360$^{\circ}$.}
 \label{fig:1}
\end{figure}

\fig{fig:1} a) illustrates the concept of the two-electron streak camera. First, the 1s electron (photo-electron) absorbs the XUV attosecond pulse
with energies above the double ionization threshold. Then, as the electron leaves
the atom it can collide with the 2s electron and transfer some of its energy, resulting in the
simultaneous ejection of both electrons.  
The intra-atomic collision is typically characterized by the asymptotic inter-electronic angle of escape, $\mathrm{\theta_{12}^{\infty}}$, that can be observed by experiment \cite{Knapp2004PRL}.  To time-resolve the two-electron collision dynamics we streak $\mathrm{\theta_{12}^{\infty}}$ by adding a weak IR laser field polarized along the z axis, $\mathrm{\vec F^{IR}(t)=F_{0}^{IR}f(t)\cos(\omega_{IR} t +\phi) \hat z}$, where $\mathrm{\phi}$ is the phase between the IR field and the XUV pulse and $\mathrm{f(t)}$ is the pulse envelope \cite{Emmanouilidou2010NJP}.
 We choose $\mathrm{\omega_{IR}= 0.0285}$ au (1600 nm) and $\mathrm{F_{0}^{IR} = 0.007}$  a.u. ($\mathrm{< 1.8 \times10^{12}}$ $\mathrm{W/cm^2}$) so that the streaking field does not alter  the attosecond collision significantly, but still has an observable effect on $\mathrm{\theta_{12}}$.  Here, $\mathrm{F_{0}^{IR}=0.007}$ a.u. is chosen to efficiently streak excess energies from 10 eV to 60 eV. The IR laser field splits $\mathrm{\theta_{12}(\phi)}$ in two branches with the lower/upper branch corresponding to launching of the photo-electron along the $\pm\mathrm{\hat{z}}$ direction, see \fig{fig:1} a).

Next, we describe how we model the XUV attosecond pulse and how its spectral intensity needs to be reflected in the weight of trajectories corresponding to different excess energies. The electric field of the XUV-pulse is of the form: 
\begin{equation}
\vec{ F}^{XUV}(t)=F_{0}^{XUV}e^{-t^2/4\sigma^2}\cos({\omega}_0^{XUV}t)\hat{z}
\label{eq:1}
\end{equation}

\noindent with $\mathrm{\sigma}$ the standard deviation of the temporal intensity envelope I(t).   For the current calculations, the spectral intensity $\mathrm{\tilde{I}(\omega)}$ of the  XUV-pulse  has a FWHM of 75 eV, centered at $\omega_0^{XUV}=120$ eV, see \fig{fig:1} b). The temporal intensity envelope $\mathrm{I(t)}$ of the transform limited pulse has a FWHM of 1a.u., see inset of \fig{fig:1} b).   In what follows we focus on the effect the large energy bandwidth of the XUV-pulse has on the streaking process and we neglect the effect of the finite FWHM of I(t). The uncertainty of the time of photo-absorption will be taken into account after the application of the streak camera algorithm as an uncertainty in the retrieved delay-times.

Using first order perturbation theory we compute the photo-absorption probability to transition from the initial ground state of He$^*$(1s2s) to the final state of double electron escape \cite{Brandsen}:

\begin{equation}
W_{i\rightarrow f} \propto \frac{1}{ \omega}\sigma^{++}(\omega)\tilde{I}(\omega)
\label{eq:2}
\end{equation}
with the cross section for double ionization $\mathrm{\sigma^{++}(\omega)}$ given by  $\mathrm{\sigma_{abs}(\omega) P^{++}(\omega)}$. $\mathrm{\sigma_{abs}}$ is  the cross-section for photo-absorption  which we calculate in the single electron approximation assuming that  the 1s electron absorbs the photon \cite{Bethe}. $\mathrm{P^{++}(\omega)}$ is the probability for double ionization obtained through our classical simulation \cite{Emmanouilidou2007PRA1, Emmanouilidou2008PRL, Emmanouilidou2010NJP}. Finally, we weight each classical trajectory for a given  photon energy $\mathrm{\omega}$ by the factor $\mathrm{\sigma_{abs}(\omega)\tilde{I}(\omega)/\omega}$.

Our goal is to retrieve the delay between photo-absorption and ionization of both electrons. Since the delay depends on the sharing of the final energy among the two electrons [10], we will consider in the following only symmetric energy sharing of $\mathrm{\epsilon<0.14}$. The delay times for the most symmetric energy sharing correspond roughly to the time of minimum approach of the two electrons, i.e. to the collision time.  Here, we have defined the energy sharing $\mathrm{\epsilon = (\epsilon_1-\epsilon_2)/(\epsilon_1 + \epsilon_2)}$, as the dimensionless asymmetry parameter between the observable (final) kinetic energies $\mathrm{\epsilon_1}$ and $\mathrm{\epsilon_2}$ of the two electrons. The analysis of different energy sharings as described in \cite{Emmanouilidou2011NJP} can be applied to the following analysis without any restrictions. In \fig{fig:1} c) we plot, a FROG-like trace for two-electron ejection, the observable total electron energy as a function of $\mathrm{\phi}$. In what follows, we describe how we extract from \fig{fig:1} c) the delay times of the intra-atomic two-electron collisions for different triggering excess energies.  

\begin{figure}
  \includegraphics[width=0.4\textwidth]{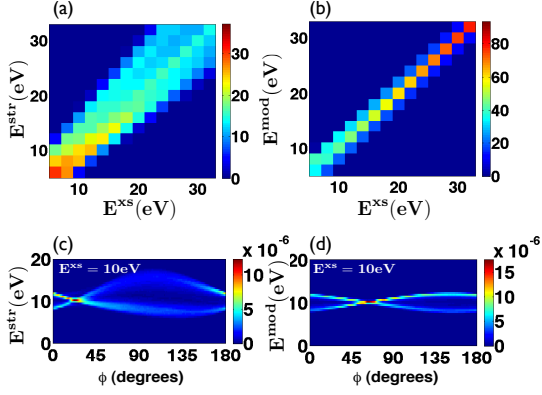}           
 \caption{Correlation map of the excess energy $\mathrm{E^{xs}}$  (a) with the observable electron energy $\mathrm{E^{str}}$ and (b) with the ``modified"  electron energy $\mathrm{E^{mod}}$. The color scale in a) and b) is such that the sum of  $\mathrm{E^{xs}}$ contributing  to a certain $\mathrm{E^{str}}$  is normalized to 100. (c) Streaked electron energy and (d) ``modified" electron energy as a function of  $\mathrm{\phi}$ for $\mathrm{E^{xs}=10}$ eV excess energy. }
  \label{fig:2}
\end{figure}

We first study the effect  of the large energy bandwidth of the XUV-pulse on streaking the two-electron collision dynamics. 
In \fig{fig:2} a) we plot the correlation map of the excess energy of the XUV-photon and the observable sum energy $\mathrm{E^{str}}$ of both electrons in the presence of the streaking IR field.
 We see that  a large range of excess energies maps to the same streaked total electron energy. Hence, the final electronic state does not correspond unambiguously to  the triggering photon energy. For instance, the 20 eV streaked energy maps to excess energies ranging from 12 eV to 26 eV. The reason for the weak correlation between the streaked and the excess energy becomes clear in  \fig{fig:2} c) for 10 eV excess energy:  the streaked energy changes significantly with $\mathrm{\phi}$.

To retrieve the excess energy from the final electronic state with improved accuracy,  we introduce a ``modified" total electron energy, where the effect of the streaking IR field is reduced.
Therefore, we define a ``modified" electron momentum $\mathrm{p_i^{mod}}$
 by subtracting the momentum change $\mathrm{\Delta p_{IR}}$ due to the streaking IR field,  i.e.
\begin{equation}
{\bf p}_{i}^{mod} = {\bf p}_{i}^ {str}- \Delta {p}_{IR}\hat{z},
\label{eq:11}
\end{equation}
where the index $\mathrm{i = 1,2}$ labels the two electrons.   The change in momentum due to the streaking field  (neglecting the Coulomb potential) is given by  
\begin{equation}
\Delta {p}_{IR} \approx \frac{F^{IR}_{0}}{\omega_{IR}}\sin(\Delta \phi +\phi) .
\label{eq:22}
\end{equation}
Here, the shift  $\mathrm{\Delta \phi = \omega_{IR}  t_{delay}}$ with respect to the maximum of the vector potential of the IR field, $\mathrm{A_{IR}(\phi)}$, (\fig{fig:1} a) is due to the delayed electron emission \cite{Emmanouilidou2010NJP}. Since we want to retrieve $\mathrm{ t_{delay}}$ we set $\mathrm{\Delta \phi = 0}$ when computing the ``modified" electron momentum $\mathrm{p_{i}^{mod}}$. Hence, $\mathrm{\Delta p_{IR}\approx \frac{F_0^{IR}}{\omega_{IR}}\sin{\phi}}$. 
  Thus, the  ``modified" energy $\mathrm{E^{mod}}$ is given by
 \begin{equation}
  \sum_{i=1,2} { \frac{((p^{str}_{x,i})^{2}+(p^{str}_{y,i})^2 )}{2}}+\sum_{i=1,2}\frac{(p^{mod}_{z,i})^2}{2}=E^{mod}
 \label{eq:44}
 \end{equation}

\fig{fig:2} d) shows that  the ``modified" electron energy varies significantly less with $\mathrm{\phi}$   compared to the unmodified, observable energy $\mathrm{E^{str}}$ (\fig{fig:2} c). Consequently, $\mathrm{E^{mod}}$ is strongly correlated with the excess energy, see  \fig{fig:2} b).
The improved correlation at higher excess energies is likely due to the faster collision, i.e. the approximation $\Delta \phi \approx 0$ is better at higher excess energies.

\begin{figure}
  \includegraphics[width=0.5\textwidth]{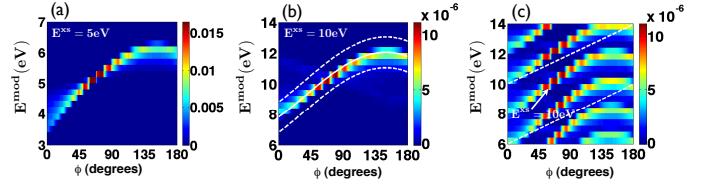}           
 \caption{(a) $\mathrm{E^{mod}}$ for one electron as a function of $\mathrm{\phi}$ for 5 eV excess energy. (b) $\mathrm{E^{mod}}$ for two electrons as a function of  $\mathrm{\phi}$ for 10 eV excess energy; white solid line shows the average of the distribution $\mathrm{E^{mod}}$  in a) times two. c) $\mathrm{E^{mod}}$ for two electrons as a function of $\mathrm{\phi}$ for excess energies between 4 and 14 eV. The white dashed lines enclose the doubly ionizing events with $\mathrm{\mathcal{E}^{mod}}=10$ eV.   }
  \label{fig:3}
\end{figure}

\begin{figure}
  \includegraphics[width=0.5\textwidth]{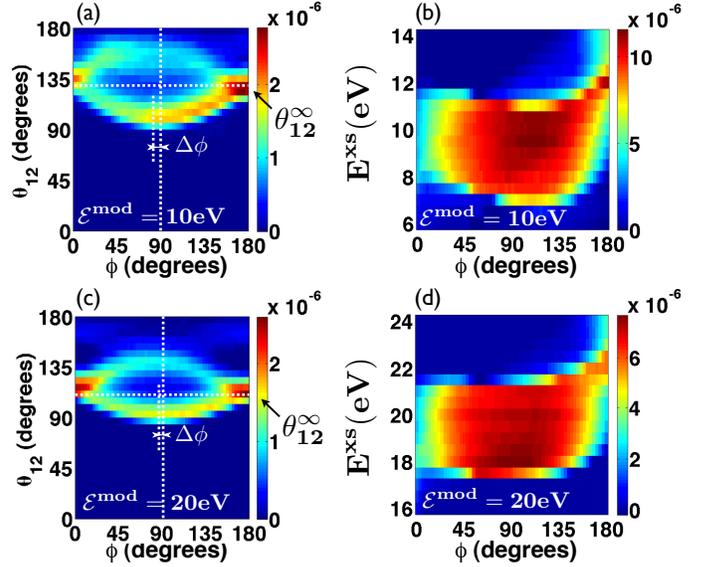}           
 \caption{$\theta_{12}$ as a function of $\phi$ for ``modified" energies around 10eV (a) and 20 eV (c) in the presence of the XUV
and IR field. $\Delta \phi$ is the shift of the maximum of the vector
potential of the IR field, corresponding to a maximum of the split of $\theta_{12}$ as a function of $\phi$.  (b) Excess energies as a function of $\phi$ that contribute to the ``modified"  energy around 10 eV enclosed by the white dashed lines in \fig{fig:3} c) and similarly (not shown) for the ``modified" energy centered around 20 eV.}
  \label{fig:4}
\end{figure}

\begin{figure}[h]
  \includegraphics[width=0.5\textwidth]{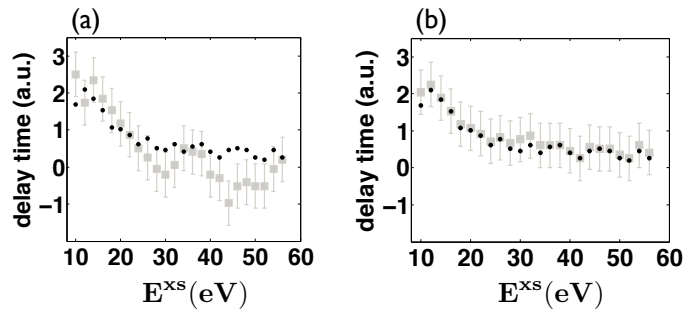}           
 \caption{ $\square$ the collision times for ``modified" electron energies $\mathrm{\mathcal{E}^{mod}}$ from 10 eV to 56 eV; $\bullet$ the collision times for excess energies $\mathrm{E^{xs}}$ ranging from 10 eV to 56 eV. Delay time was retrieved by a) method 1 and b) method 2. The error bars show the uncertainty in the delay times of 0.6 a.u. since we change $\mathrm{\phi}$ every one degree.  }
 \label{fig:5}
\end{figure}

 We next explain why at $\mathrm{\phi=0^{\circ}/180^{\circ}}$ the ``modified" electron energy is smaller/larger than the corresponding excess energy for photo-electrons ejected in the +$\mathrm{\hat{z}}$ direction (it is the other way around for photo-electrons ejected in the -$\mathrm{\hat{z}}$ direction). At $\mathrm{\phi=0^{\circ}}$ each electron experiences  a force from the IR field  in the direction opposite to the electron's direction of escape. As a result, each electron slows down and escapes with an energy smaller than the electron's final energy in the absence of the IR field. In contrast, at $ \mathrm{\phi=180^{\circ}}$
 each electron experiences a force from the IR field in the same direction as the electron's direction of escape. As a result, each electron escapes with an energy larger than the electron's final energy in the absence of the IR field. To verify
 that the overall change of the ``modified" total electron energy with $\mathrm{\phi}$ is a one-electron effect, we run our simulation in the presence of the XUV plus IR field only for the photo-electron (the 2s electron is absent). 
 Since for the two electron case we only consider equal energy sharing, we compare the two electron case for a certain excess energy with the one electron case for half that excess energy.
 Indeed, multiplying by two the distribution of the ``modified" one-electron energy as a function of $\mathrm{\phi}$ for an excess energy of, for example, 5 eV (\fig{fig:3} a) and taking the average we find that there is excellent agreement with the two electron ``modified" electron energy at 10 eV excess energy as a function of $\mathrm{\phi}$, see \fig{fig:3} b). Note that  in \fig{fig:3}  and in what follows (\fig{fig:4} b) and d) and \fig{fig:5}) we focus on double ionization events where the photo-electron is ejected in the +$\mathrm{\hat{z}}$ direction. 

 The increase of the ``modified" electron energy with $\mathrm{\phi}$ for each excess energy $\mathrm{E^{xs}}$ when the photo-electron is ejected in the +$\mathrm{\hat{z}}$ direction, forms the basis for the simple algorithm we develop to restrict the residual variation between excess energy and ``modified" electron energy, i.e. the range of ``modified" electron energies that pertain to a certain excess energy for each value of $\mathrm{\phi}$. 
If our algorithm is dictated mainly by experimentally accessible observables, we compute the collision time corresponding to $\mathrm{E^{xs}}$ by selecting the doubly ionizing events whose ``modified" electron energy changes from  $\mathrm{[E^{xs}-\Delta E/2, E^{xs}]}$ eV for $\mathrm{\phi=0^{\circ}}$ to $\mathrm{[E^{xs}+\Delta E/2,E^{xs}]}$ for $\mathrm{\phi=180^{\circ}}$. Choosing $\mathrm{\Delta E}$ to be roughly 8 eV for all excess energies (method 1), the thus selected doubly ionizing events have excess energies close to the desired $\mathrm{E^{xs}}$ for all $\mathcal{\phi}$'s, and we label them by $\mathrm{\mathcal{E}^{mod}}$.  Indeed, the double ionization events with  $\mathrm{\mathcal{E}^{mod}=10}$ eV,  illustrated with the white dashed lines in \fig{fig:3} c),  are the events corresponding to excess energies close to $\mathrm{E^{xs}=10 eV}$ for all $\mathrm{\phi}$'s, see \fig{fig:4} b) and similarly for $\mathrm{\mathcal{E}^{mod}=20 eV}$  see \fig{fig:4} d). We then determine the collision time for $\mathrm{\mathcal{E}^{mod}=10 eV/20 eV}$ in \fig{fig:4} a) and c) by extracting $\mathrm{\Delta \phi}$ from the lower branch of the inter-electronic angle of escape $\mathrm{\theta_{12}(\phi)}$  \cite{Emmanouilidou2010NJP,Emmanouilidou2011NJP}.  We find  that $\mathrm{\Delta \phi}$ is 4.1$^{\circ}$/1.9$^{\circ}$ corresponding to a collision time of 2.5~a.u./1.2~a.u. for $\mathrm{\mathcal{E}^{mod}=10 eV/20 eV}$, respectively.

We note that the variation in collision time with excess energy suggests that the accuracy of $\mathrm{\Delta \mathrm{\phi}}$ depends critically on the resolution in $\theta_{12}$. In order to increase the robustness of the retrieval algorithm we determine $\Delta \phi$ for a range of bin sizes $d\theta_{12} = 4^{\circ}-9^{\circ}$. We define the average of $\Delta \phi(d\theta_{12})$ as the collision phase or collision time at a given excess energy or $\mathrm{\mathcal{E}^{mod}}$. 

In \fig{fig:5} a) we show the collision time for $\mathrm{\mathcal{E}^{mod}}$ ranging from 10~eV to 56~eV excess energies. We find that the collision time decreases with increasing excess energy, as expected. Moreover, we find that the agreement between the collision times for a certain  $\mathrm{\mathcal{E}^{mod}}$ and the corresponding  $\mathrm{E}^{xs}$ is best for smaller excess energies. The reason is that we compute the delay times corresponding to a certain  $\mathrm{\mathcal{E}^{mod}}$ using $\mathrm{\Delta E}\approx 8$~eV independent of the excess energy. This choice of $\mathrm{\Delta E}$ describes best the rate of increase of the ``modified" electron energy with $\mathrm{\phi}$ for smaller excess energies.  However, $\mathrm{\Delta E}$ decreases with increasing excess energy. As a result, the agreement is worse for higher excess energies.

\begin{figure}[h]
  \includegraphics[width=0.4\textwidth]{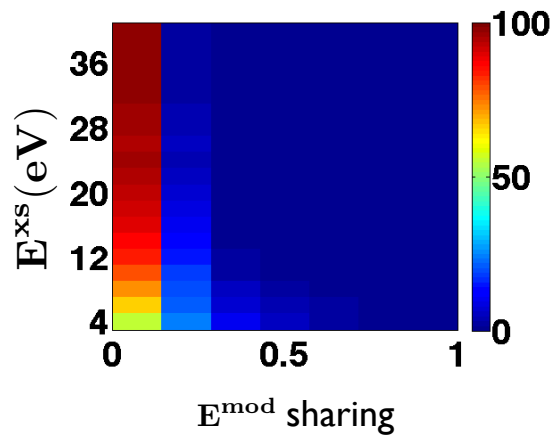}           
 \caption{ Correlation map of the excess energy with the ``modified" energy sharing for the doubly ionizing events with equal energy sharing in the absence of the IR field. The color scale is such that the sum of  $\mathrm{E^{mod}} $ energy sharings contributing  to the equal energy sharing double ionization events for a certain $\mathrm{E^{xs}}$  is normalized to 100. }
 \label{fig:6}
\end{figure}
Accounting for the fact that $\mathrm{\Delta E}$ changes with excess energy (method 2) we obtain a much better agreement between the two sets of collision time, see \fig{fig:5} b). One way to do so is by labelling as $\mathrm{\mathcal{E}^{mod}}$ the doubly ionizing trajectories with $\mathrm{E^{mod}}$ within $\pm$ 1~eV of twice the average $\mathrm{E^{mod}}$ for the one electron problem, see \fig{fig:3} a) and b). However, in this latter algorithm we use as input the computed $\mathrm{E^{mod}}$ for one electron for each excess energy while in the previous paragraph we only  use experimentally accessible data.  
Finally let us note that both in \fig{fig:5} a) and b) the collision times are computed for equal ``modified" energy sharing. Similarly to the total ``modified" electron energy which is strongly correlated to the excess energy (\fig{fig:2} b) the ``modified" equal energy sharing is strongly correlated to the equal energy sharing in the absence of the IR field (\fig{fig:6}); it is the delay times corresponding to the latter energy sharings that we aim to streak.

Concluding, we have demonstrated that the two-electron streak camera can be experimentally realized. We obtain a complete picture of the single-photon double ionization process by computing the  intra-atomic two-electron collision times for different excess energies.  While our study has been performed in the context of two electron escape in atoms it opens the way for  time-resolving collision dynamics during multi-electron escape in atomic and molecular systems.

\end{document}